\documentclass[a4paper,11pt]{article}
\usepackage[dvips]{graphicx}
\usepackage{amssymb}
\usepackage{amsmath}

\usepackage{cite}

\begin{document}

\title{Localized Fermions on Superconducting Domain Walls and Extended Supersymmetry with non-trivial Topological Charges}
\author{
V.K. Oikonomou$^{1,}$$^{2}$\,\thanks{voiko@physics.auth.gr}\\
$^1$Department of Mechanical Engineering, Technological Education Institute of Serres\\
62124 Serres, Greece \\
$^2$Department of Theoretical Physics, Aristotle University of Thessaloniki,\\
54124 Thessaloniki, Greece
} \maketitle

\begin{abstract}
In this letter we demonstrate that the fermionic zero modes on a superconducting domain wall can be associated to an one dimensional $N=6$ supersymmetry that contains non-trivial topological charges. In addition, the system also possesses three distinct $N=4$ supersymmetries with non-trivial topological charges and we also study some duality transformations of the supersymmetric algebras.
\end{abstract}

\section*{Introduction and Motivation}

Domain walls, cosmic strings and monopoles, are generated in realistic grand unified theories beyond the Standard Model \cite{vilenkin,lazaridesvasiko}, with the least problematic defect among these being cosmic strings with high energy symmetry breaking scale, with regards to cosmology. Monopoles and domain walls on the other hand are responsible for many cosmological inconsistencies. Nevertheless, topologically unstable domain walls can be phenomenologically acceptable for various reasons \cite{lazaridesvasiko}. In fact, topologically unstable domain walls can be plausible, since a low tension domain wall network in the universe can be responsible for dark energy, thus providing us with a non-exotic alternative to existing dark energy models related to $f(R)$ theories of gravity (for an important stream of papers on $f(R)$ gravity dark energy models see \cite{odintsovgravity} and references therein). In addition, domain walls bounded by strings are topologically unstable and can disappear before dominating the expansion of the universe, but however these are locally stable and lose energy through their interaction with the surrounding medium. 

Domain walls are topological defects created by the spontaneous symmetry breaking of a discrete symmetry of a gauge field theory containing Higgs scalars, with the discrete broken symmetry not being part of the gauge symmetry of the gauge field theory. As a consequence of the spontaneous symmetry breaking, the total vacuum manifold consists of many distinct vacuum states, with the scalar field related to the spontaneously broken symmetry taking one of these distinct vacuum states \cite{lazaridesvasiko}. The domain walls separate the different vacuum regions, with the scalar field interpolating between these distinct vacuum states. In this paper we shall be interested in superconducting domain walls \cite{lazaridesvasiko}. A superconducting domain wall is a domain wall with localized fermionic zero modes which can move along the wall with the wall acquiring charge and electric current. 

In a previous paper \cite{oikonomoudomain} we demonstrated that the fermionic zero modes localized on topologically unstable domain walls bounded by strings in a grand unified theory theoretical framework, can be connected to three independent $N=2$, $d=1$ supersymmetric quantum mechanics algebras (for reviews and important papers on supersymmetry see \cite{reviewsusyqm,diffgeomsusyduyalities,various,susyqminquantumsystems,plu1,plu2,susyqmscatter,susybreaking,ivanov,extendedsusy,witten1}). The fermions we took into account were up quarks, down quarks and also charged leptons. An important question we raised and answered for a particular case, was whether these supersymmetries can be combined in some way to form a higher order extended supersymmetry, and as we explicitly showed, this can be done in the particular case that the couplings of the charged lepton and the down quark become equal. This can occur at some coupling unification scale $M_s$, related to the specific grand unified theory under study. The extended supersymmetric structure was that of an $N=4$, $d=1$ supersymmetry with central charges. This phenomenon occurred due to the fact that the domain wall kink solution couples in the same way to the down-quarks and to the charged leptons. In this paper we shall present a much more general supersymmetric structure of the up-down quark and charged lepton system of fermions on the domain wall, since the extended supersymmetric structure in the present paper occurs regardless the coupling unification of the charged lepton and down-quark system. In fact, it contains all the supercharges corresponding to all the aforementioned fermions. As we shall demonstrate in detail, we can form an one dimensional $N=6$ supersymmetry consisting of three supercharges and additionally three distinct $N=4$, $d=1$ supersymmetries. Both the supersymmetries we found contain non-trivial topological charges that do not commute with all the operators of the algebra. We shall also study various dualities that exist in the supersymmetric algebras.

Supersymmetric quantum mechanics \cite{reviewsusyqm} (abbreviated to SUSY QM hereafter) which was originally introduced to model supersymmetry breaking in quantum field theory, has developed to be an independent research field, with various applications in many research areas. Hilbert space properties corresponding to SUSY QM systems and in addition various applications to quantum mechanical systems were presented in \cite{diffgeomsusyduyalities} and \cite{susyqminquantumsystems,plu1,plu2} respectively. Applications of SUSY QM to scattering related phenomena can be found in \cite{susyqmscatter} and also, various features of supersymmetry breaking were presented in \cite{susybreaking}. Higher $N$-extended one dimensional supersymmetries and their connection to harmonic superspaces or gravity were studied in \cite{extendedsusy,ivanov}. For applications of supersymmetry in quantum field theory see \cite{witten1,odi1,odi2,odi3}. However, it is known that the representations of higher N-extended supersymmetry in dimension four are difficult to find, therefore dimensionally reduced SUSY QM models can be a great simplification of the problem and can provide us with important insights regarding these issues. In this letter we actually found an enhanced supersymmetry of the field theory but not of the S-matrix of the model and we believe that this enhanced symmetry containing non-trivial supercharges, can be an indication of a more involved symmetry yet to be found.

This paper is organized as follows: In section 1 we briefly review all the necessary information concerning fermions on a domain wall and their connection to the three distinct $N=2$, $d=1$ supersymmetries, following reference \cite{oikonomoudomain}. In addition we shall also present the theoretical framework of the domain wall model we shall use. In section 2, we present all the details concerning the enhanced underlying $N=6$, $d=1$ supersymmetric structure of the fermionic system . In section 3 we present the three different $N=4$, $d=1$ supersymmetries that the system possesses and in section 4 we study some dualities that the supersymmetric algebras have. The conclusions follow in the end of the paper.

\section{A Brief Review of the Theoretical Framework-Superconducting Domain Walls and $N=2$, $d=1$ Supersymmetry}

In order to make the article self contained, we briefly present the essentials of the grand unified domain wall model and also the connection of the localized fermions with $N=2$, $d=1$ supersymmetry. For details on the issues that will be presented, see \cite{lazaridesvasiko} regarding the domain wall model and also \cite{oikonomoudomain} for the connection of the localized fermions with the one dimensional supersymmetries. Consider a symmetry breaking pattern of a grand unified theory with an $SO(10)$ gauge group, for which grand unified model we have domain walls bounded by cosmic strings. The main focus in this article will be in the $126$-Higgs representation breaking pattern, which actually is:
\begin{equation}\label{symmetrybreakpatt}
Spin(10)\xrightarrow{54,M_x}H_1\xrightarrow{126,M_R}H_2\xrightarrow{10,M_w}SU(3)_c\times U(1)_{em}
\end{equation}
Note that the actual grand unification symmetry of the system is not the $SO(10)$ group, but the $Spin(10)$, since the fermions belong to the $16$-representation, that is, the fundamental spinor representation of $SO(10)$. In relation (\ref{symmetrybreakpatt}), the mass scales $M_x,M_R$ are of the order $M_x\sim 10^{15}$GeV and $M_R\sim 10^{13}$GeV. Moreover, the subgroups $H_1$ and $H_2$ are equal to:
\begin{align}\label{subgroupsbrek}
& H_1\sim Spin(6)\times Spin(4) \\ \notag & H_2\sim SU(3)_c\times SU(2)_L\times U(1)
\end{align}
with the group $H_1$ being isomorphic to the Pati-Salam subgroup, that is, $H_1\sim SU(4)\times SU(2)_L\times SU(2)_R$. Owing to the fact that the subgroup $H_1$ is disconnected, the fundamental homotopy group is $\pi (H_1)=Z_2$, with this $Z_2$ symmetry being generated by the disconnected piece of $H_1$, which we denote $C$. Additionally, note that $H_1=H_0'\times C$, with the subgroup $H_0'$ being equal to:
\begin{equation}\label{hosubgroup}
H_0'=(Spin(6)\times Spin(4))/Z_2
\end{equation}
When the symmetry of the system is $H_1$, topologically stable strings occur and at the end of the second symmetry breaking stage, the discrete symmetry $C$ breaking occurs. As a consequence of this symmetry breaking, domain walls appear which separate regions with opposite values of the symmetry $C$. The domain walls ends are on the $H_1$-phase strings, with the domain walls being unstable, due to the possible decay into holes bounded by string loops. We denote the scalar field that will give rise to domain walls by $\phi$, and it can take two vacuum expectation values, namely $\langle \phi \rangle=\phi_u$ and $\langle \phi \rangle=\phi_d$. The vacuum expectation values $-\langle \phi \rangle$ and $\langle \phi \rangle$, can be connected with the aid of a kink configuration, when moving along the domain wall. On the domain wall there exist fermionic localized degrees of freedom, which are massless and it is exactly the existence of these massless localized zero modes that makes the domain wall superconducting \cite{lazaridesvasiko}. 

Regarding the localized fermions on the wall, there exist both left handed fields $\psi_i=(u_L,d_L,e_L)$ and right handed fields $\chi_i=(u_R,d_R,e_R)$, with $(e,u,d)$ denoting the electron field and up and down quarks respectively. Notice that $\phi_d$ couples to the down quark and the charged leptons, while $\phi_u$ couples to the up-quarks only. This fact plays some role on the transformation properties of the underlying supersymmetric algebra, as we shall demonstrate in a latter section. Assuming an infinite wall in the $x-z$ plane and in addition that $A_i=0$, the $y$-direction transverse fermionic zero modes satisfy the following equations of motion \cite{lazaridesvasiko}:
\begin{align}\label{fer1}
 -&i\sigma^2\partial_y\psi_i(y)-g_i\phi_i^{kink}(y)\chi_i(y)=0
\\ \notag & i\sigma^2\partial_y\chi_i(y)-g_i\phi_i^{kink}(y)\psi_i(y)=0
\end{align}
with $i=u,d,e$. As was noticed in \cite{lazaridesvasiko}, a real kink solution $\phi_i(y)=\phi_i^{kink}(y)$ can always be found, therefore in the following we shall assume that $\phi_i^{kink}(y)$ is real and in addition that $g_i$ is a real coupling which takes the values $g_i=(g_u,g_d,g_e)$ when $\phi (y)$ is coupled to the up quark, the down quark, and the charged lepton respectively. Note that $\phi_e^{kink}(y)=\phi_d^{kink}(y)$ since the kink couples in the same way to the down quark and lepton sector and we shall use only $\phi_d^{kink}(y)$ hereafter. The above equations (\ref{fer1}) admit localized fermionic solutions, which are:
\begin{align}\label{soleqnts}
& \psi_i(y)=c_ie^{-g_i\int_0^y\phi_i^{kink}(y)\mathrm{d}y},{\,}{\,}{\,}\chi_i(y)=c_ii\sigma^2e^{-g_i\int_0^y\phi_i^{kink}(y)\mathrm{d}y}
\end{align}
where $i=u,d,e$. Actually, the existence of these localized transverse zero modes is what makes the domain wall superconducting. We introduce the operator $D_i$:
\begin{equation}\label{susyqmrn567m}
\mathcal{D}_{i}=\left(%
\begin{array}{cc}
-i\sigma^2\partial_y & g_i\phi_i^{kink}(y)
 \\ g_u\phi_i^{kink}(y) & i\sigma^2\partial_y\\
\end{array}%
\right)
\end{equation}
which actually describes the equations of motion (\ref{fer1}) and its zero modes are the solutions (\ref{soleqnts}). As was demonstrated in \cite{oikonomoudomain}, using this operator we can construct three unbroken underlying $N=2$, $d=1$ supersymmetries, one for each fermion pair $(\psi_i(y),\chi_i(y))$, with $i=u,d,e$. The supercharges of the algebras, denoted $\mathcal{Q}_d$, $\mathcal{Q}_e$ and $\mathcal{Q}_u$ are equal to:
\begin{equation}\label{s7supcghagemix}
\mathcal{Q}_{d}=\bigg{(}\begin{array}{ccc}
  0 & \mathcal{D}_{d} \\
  0 & 0  \\
\end{array}\bigg{)},{\,}{\,}{\,}\mathcal{Q}_{e}=\bigg{(}\begin{array}{ccc}
  0 &  \mathcal{D}_{e} \\
 0 & 0  \\
\end{array}\bigg{)},{\,}{\,}{\,}\mathcal{Q}_{u}=\bigg{(}\begin{array}{ccc}
  0 & \mathcal{D}_{u} \\
  0 & 0  \\
\end{array}\bigg{)}
\end{equation}
In addition, the quantum Hamiltonians of the three superalgebras are:
\begin{equation}\label{s11fgghhf}
\mathcal{H}_{d}=\bigg{(}\begin{array}{ccc}
 \mathcal{D}_{d}\mathcal{D}_{d}^{\dag} & 0 \\
  0 & \mathcal{D}_{d}^{\dag}\mathcal{D}_{d}  \\
\end{array}\bigg{)},{\,}{\,}{\,}\mathcal{H}_{e}=\bigg{(}\begin{array}{ccc}
 \mathcal{D}_{e}\mathcal{D}_{e}^{\dag} & 0 \\
  0 & \mathcal{D}_{e}^{\dag}\mathcal{D}_{e}  \\
\end{array}\bigg{)},{\,}{\,}{\,}\mathcal{H}_{u}=\bigg{(}\begin{array}{ccc}
 \mathcal{D}_{u}\mathcal{D}_{u}^{\dag} & 0 \\
  0 & \mathcal{D}_{u}^{\dag}\mathcal{D}_{u}  \\
\end{array}\bigg{)}
\end{equation}
In the following sections, we shall make extensive use of the notation and operators we introduced in this section.

\section{$N=6$, $d=1$ Extended Supersymmetry with non-trivial Topological Charges}

As was demonstrated in reference \cite{oikonomoudomain}, the two of the three $N=2$, $d=1$ algebras that the fermionic system possesses, combine under certain circumstances to form an extended $N=4$, $d=1$ supersymmetric algebra with non-trivial central charges. Particularly, owing to the fact that the kink couples in the same way to the down quark and lepton sector, when the coupling constants $g_e$ and $g_d$ become equal, then the system of fermions composed by the down quark and charged lepton possesses the extended $N=4$ supersymmetric structure. This result however is not so general since it depends on the equality of the coupling constants. In this section we shall present a much more general case in which the $N=2$ superalgebras combine to form higher extended supersymmetric structures, a result that is independent of the specifics of the model. As we shall demonstrate, the three superalgebras combine to form an $N=6$, $d=1$ supersymmetry with non-trivial topological charges (for important and useful discussions on topological charges, consult \cite{wittentplc,fayet,topologicalcharges,topologicalcharges1}). Note that in the present case we have non-trivial topological charges which, in contrast to central charges of the previous case, do not commute with all the operators of the superalgebra. We shall discuss this issue and what it may imply for underlying algebraic structures, in the end of this section. 

In order to reveal the underlying extended supersymmetry, consider the supercharges appearing in equation (\ref{s7supcghagemix}) and we compute the following commutation and anticommutation relations:
\begin{align}\label{n4algbe1sjdjfgeneral}
&\{{{\mathcal{Q}}_{u}},{{\mathcal{Q}}^{\dag}_{u}}\}=2\mathcal{H}+\mathcal{Z}_{uu},
{\,}\{{{\mathcal{Q}}_{e}},{{\mathcal{Q}}^{\dag}_{e}}\}=2\mathcal{H}+\mathcal{Z}_{ee}
,{\,}\{{{\mathcal{Q}}_{e}},{{\mathcal{Q}}_{e}}\}=0,
\\ \notag & {\,} \{{{\mathcal{Q}}_{e}},{{\mathcal{Q}}_{ u}}^{\dag}\}=\mathcal{Z}_{e u},{\,}\{{{\mathcal{Q}}_{u}},{{\mathcal{Q}}^{\dag}_{e}}\}=\mathcal{Z}_{u e},{\,}\{{{\mathcal{Q}}^{\dag}_{u}},{{\mathcal{Q}}^{\dag}_{u}}\}=0\\ \notag
&\{{{\mathcal{Q}}^{\dag}_{e}},{{\mathcal{Q}}^{\dag}_{e}}\}=0,{\,}\{{{\mathcal{Q}}^{\dag}_{e}},{{\mathcal{Q}}^{\dag}_{u}}\}=0,{\,}\{{{\mathcal{Q}}_{e}},{{\mathcal{Q}}_{u}}\}=0{\,}\\
\notag
&[{{\mathcal{Q}}_{u}},{{\mathcal{Q}}_{e}}]=0,[{{\mathcal{Q}}^{\dag}_{e}},{{\mathcal{Q}}^{\dag}_{u}}]=0,{\,}[{{\mathcal{Q}}_{u}},{{\mathcal{Q}}_{u}}]=0,{\,}[{{\mathcal{Q}}^{\dag}_{u}},{{\mathcal{Q}}^{\dag}_{u}}]=0
\\ \notag &\{{{\mathcal{Q}}_{e}},{{\mathcal{Q}}^{\dag}_{e}}\}=2\mathcal{H}+\mathcal{Z}_{ee},
{\,}\{{{\mathcal{Q}}_{d}},{{\mathcal{Q}}^{\dag}_{d}}\}=2\mathcal{H}+\mathcal{Z}_{dd}
,{\,}\{{{\mathcal{Q}}_{d}},{{\mathcal{Q}}_{d}}\}=0,
\\ \notag & {\,} \{{{\mathcal{Q}}_{d}},{{\mathcal{Q}}_{ e}}^{\dag}\}=\mathcal{Z}_{d e},{\,}\{{{\mathcal{Q}}_{e}},{{\mathcal{Q}}^{\dag}_{d}}\}=\mathcal{Z}_{e d},{\,}\\ \notag
&\{{{\mathcal{Q}}^{\dag}_{d}},{{\mathcal{Q}}^{\dag}_{d}}\}=0,{\,}\{{{\mathcal{Q}}^{\dag}_{d}},{{\mathcal{Q}}^{\dag}_{e}}\}=0,{\,}\{{{\mathcal{Q}}_{d}},{{\mathcal{Q}}_{e}}\}=0{\,}\\
\notag
&[{{\mathcal{Q}}_{e}},{{\mathcal{Q}}_{d}}]=0,[{{\mathcal{Q}}^{\dag}_{d}},{{\mathcal{Q}}^{\dag}_{e}}]=0,{\,}[{{\mathcal{Q}}_{e}},{{\mathcal{Q}}_{e}}]=0,{\,}[{{\mathcal{Q}}^{\dag}_{e}},{{\mathcal{Q}}^{\dag}_{e}}]=0
\\ \notag &\{{{\mathcal{Q}}_{u}},{{\mathcal{Q}}_{ u}}\}=0, {\,} \{{{\mathcal{Q}}_{d}},{{\mathcal{Q}}^{\dag}_{ u}}\}=\mathcal{Z}_{d u},{\,}\{{{\mathcal{Q}}_{u}},{{\mathcal{Q}}^{\dag}_{d}}\}=\mathcal{Z}_{u d},{\,}[{{\mathcal{Q}}_{u}},{{\mathcal{Q}}_{d}}]=0\\ \notag
&\{{{\mathcal{Q}}^{\dag}_{u}},{{\mathcal{Q}}^{\dag}_{u}}\}=0,{\,}\{{{\mathcal{Q}}^{\dag}_{d}},{{\mathcal{Q}}^{\dag}_{u}}\}=0,{\,}\{{{\mathcal{Q}}_{d}},{{\mathcal{Q}}_{u}}\}=0{\,}[{{\mathcal{Q}}^{\dag}_{d}},{{\mathcal{Q}}^{\dag}_{u}}]=0\\
\end{align}
where the operator $\mathcal{H}$ stands for:
\begin{equation}\label{newsusymat12}
\mathcal{H}=\frac{1}{2} \left ( \begin{array}{ccccc}
  \sigma_2^2\partial_y^2 & 0 & 0 & 0\\
  0 & \sigma_2^2\partial_y^2 & 0 & 0 \\
  0 & 0 & \sigma_2^2\partial_y^2 & 0 \\
  0 & 0 & 0 & \sigma_2^2\partial_y^2 \\
\end{array}\right ).
\end{equation}
In relation (\ref{n4algbe1sjdjfgeneral}) notice that there are many non-trivial topological charges which result from, or are equal to, anticommutators of Fermi (supercharges) charges. These topological charges are non-trivial and we shall present them in detail. We describe first the topological charges that result from anticommutation relations of the form $\{ \mathcal{Q}_i,\mathcal{Q}_i^{\dag}\}=2\mathcal{H}+\mathcal{Z}_{ii}$, with $i=u,d,e$. The general form of the topological charges $\mathcal{Z}_{ii}$ is:
\begin{equation}\label{newsusymat22}
\mathcal{Z}_{ii }=\left ( \begin{array}{ccccc}
  \mathcal{Z}^1_{ii } & 0 \\
  0 & \mathcal{Z}^2_{ii }  \\
\end{array}\right ).
\end{equation}
with the operator $\mathcal{Z}^1_{ii }$ being equal to the following matrix:
\begin{equation}\label{newsusymat32}
\mathcal{Z}^1_{ii }=\left ( \begin{array}{cc}
 g_i^2\Big{(}\phi_i^{kink}(y)\Big{)}^2 & -i\sigma_2 \partial_y \Big{(}g_i\phi_i^{kink}(y)\Big{)}-ig_i\phi_i^{kink}(y)\sigma_2\partial_y \\
 ig_i\phi_i^{kink}(y)\sigma_2\partial_y+i\sigma_2\partial_y\Big{(}g_i\phi_i^{kink}(y)\Big{)} & g_i^2\Big{(}\phi_i^{kink}(y)\Big{)}^2 \\
\end{array}\right )
\end{equation}
and correspondingly, the operator $\mathcal{Z}^2_{ii }$, is equal to the matrix:
\begin{equation}\label{newsusymat42}
\mathcal{Z}^2_{ii }=\mathcal{Z}^1_{ii }=\left ( \begin{array}{cc}
 g_i^2\Big{(}\phi_i^{kink}(y)\Big{)}^2 & -i\sigma_2 \partial_y \Big{(}g_i\phi_i^{kink}(y)\Big{)}-ig_i\phi_i^{kink}(y)\sigma_2\partial_y \\
 ig_i\phi_i^{kink}(y)\sigma_2\partial_y+i\sigma_2\partial_y\Big{(}g_i\phi_i^{kink}(y)\Big{)} & g_i^2\Big{(}\phi_i^{kink}(y)\Big{)}^2 \\
\end{array}\right )
\end{equation}
with $i=u,d,e$. The topological charges that are directly equal to the anticommutator  $\{\mathcal{Q}_i,\mathcal{Q}_{ j}^{\dag}\}=\mathcal{Z}_{ij}$, with $i\neq j$ and $i,j=u,d,e$, are equal to:
\begin{equation}\label{newsusymat52}
\mathcal{Z}_{i j }=\left ( \begin{array}{ccccc}
  \mathcal{Z}^1_{i j } & 0 \\
  0 & \mathcal{Z}^2_{i j }  \\
\end{array}\right ).
\end{equation}
with $\mathcal{Z}^1_{i j }$ being the matrix:
\begin{equation}\label{newsusymat62}
\mathcal{Z}^1_{i j }=\left ( \begin{array}{cc}
 \sigma_2^2\partial_y^2+g_ig_j\phi_j^{kink}(y)\phi_i^{kink}(y) & -i\sigma_2 \partial_y \Big{(}g_j\phi_j^{kink}(y)\Big{)}-ig_i\phi_i^{kink}(y)\sigma_2\partial_y \\
 ig_i\phi_i^{kink}(y)\sigma_2\partial_y+i\sigma_2\partial_y\Big{(}g_j\phi_j^{kink}(y)\Big{)} & g_ig_j\phi_j^{kink}(y)\phi_i^{kink}(y) +\sigma_2^2\partial_y^2\\
\end{array}\right )
\end{equation}
and also the operator $\mathcal{Z}^2_{\alpha \beta i j }$ is equal to:
\begin{equation}\label{newsusymat72}
\mathcal{Z}^2_{i j }=\left ( \begin{array}{cc}
 \sigma_2^2\partial_y^2+g_ig_j\phi_j^{kink}(y)\phi_i^{kink}(y) & i\sigma_2 \partial_y \Big{(}g_i\phi_i^{kink}(y)\Big{)}+ig_j\phi_j^{kink}(y)\sigma_2\partial_y \\
 -ig_j\phi_j^{kink}(y)\sigma_2\partial_y-i\sigma_2\partial_y\Big{(}g_i\phi_i^{kink}(y)\Big{)} & g_ig_j\phi_i^{kink}(y)\phi_j^{kink}(y) +\sigma_2^2\partial_y^2\\
\end{array}\right )
\end{equation}
The anticommutation and commutation relations (\ref{n4algbe1sjdjfgeneral}) describe an $N=6$, $d=1$ supersymmetric algebra with non-trivial topological charges. The supercharges of the algebra are $\mathcal{Q}_u,\mathcal{Q}_d,\mathcal{Q}_e$ and the topological charges are $\mathcal{Z}_{ee},\mathcal{Z}_{dd},\mathcal{Z}_{uu},\mathcal{Z}_{ud},\mathcal{Z}_{ed},\mathcal{Z}_{ue}$ and their complex conjugates. Notice that the complex conjugate topological charges can be found easily, since $\mathcal{Z}_{ij}=\mathcal{Z}_{ji}^{\dag}$. The topological charges $\mathcal{Z}_{ee},\mathcal{Z}_{dd},\mathcal{Z}_{uu}$ can be found using equation (\ref{newsusymat22}), while the rest of them, that is $\mathcal{Z}_{ud},\mathcal{Z}_{ed},\mathcal{Z}_{ue}$ can be found using (\ref{newsusymat52}).

Before closing this section, it worths discussing on the issue of topological charges and the possible underlying structures that these may imply. Topological charges occur quite often in supersymmetric algebras and were first noticed in \cite{wittentplc}, where the terminology topological charge was used for the first time. In principle, the topological charges cannot be considered as central charges \cite{fayet}, since these have non-trivial and non-zero commutation relations with the operators of the superalgebra. Furthermore, the topological charges are actually symmetries of the field theory but not of the $S$-matrix of the theory \cite{fayet}, something that applies to our case too. Possibly, the topological charges we found in this paper are not symmetries of the $S$-matrix, but indicate some additional external symmetry of the field theory describing the fermionic zero modes on the superconducting domain wall. It worths noticing that the theoretical framework used in \cite{wittentplc}, was a supersymmetric model in the presence of topological defects and as was indicated, the presence of topological charges in such frameworks are unavoidable. The interplay between topological defects and supersymmetry is particularly interesting and was also noticed in \cite{topologicalcharges}. With regard to the possible form of the underlying algebra, as was indicated in \cite{topologicalcharges1}, the existence in the algebra of a ''central'' charge that has non-trivial (non-zero) commutation relations with the rest operators of the algebra, indicates a non-linear richer supersymmetric structure. Before closing, let us note that non-commuting topological charges occur quite frequently in string theory contexts, for example in Ad$S_5\times S^5$ D-brane background with superalgebra $su(2,2|4)$ \cite{topologicalcharges2}, with the latter being maximal extension the $osp(1|32)$. The topological charges in these cases have also non trivial commutation relations with the elements of the superalgebra. 

\section{Three Distinct $N=4$, $d=1$ Supersymmetries with non-trivial Topological Charges}

Apart from the $N=6$, $d=1$ superalgebra we found, the fermionic system possesses three distinct $N=4$, $d=1$ superalgebras. These algebras contain two complex supercharges which can be composed using every combination of two supercharges, out of the three available. In this section we shall present the three different $N=4$, $d=1$ superalgebras by directly computing the corresponding commutators and anticommutators.

The first $N=4$, $d=1$ superalgebra contains the supercharges ${\mathcal{Q}}_{u}$ and $\mathcal{Q}_d$. In order to reveal the $N=4$ superalgebra, we compute the following commutators and anticommutators:
\begin{align}\label{commutatorsanticomm}
&\{{{\mathcal{Q}}_{u}},{{\mathcal{Q}}^{\dag}_{u}}\}=2\mathcal{H}+\mathcal{Z}_{uu},
{\,}\{{{\mathcal{Q}}_{d}},{{\mathcal{Q}}^{\dag}_{d}}\}=2\mathcal{H}+\mathcal{Z}_{dd}
,{\,}\{{{\mathcal{Q}}_{d}},{{\mathcal{Q}}_{d}}\}=0,
\\ \notag &\{{{\mathcal{Q}}_{u}},{{\mathcal{Q}}_{ u}}\}=0, {\,} \{{{\mathcal{Q}}_{d}},{{\mathcal{Q}}_{ u}}^{\dag}\}=\mathcal{Z}_{d u},{\,}\{{{\mathcal{Q}}_{u}},{{\mathcal{Q}}^{\dag}_{d}}\}=\mathcal{Z}_{u d},{\,}\\ \notag
&\{{{\mathcal{Q}}^{\dag}_{u}},{{\mathcal{Q}}^{\dag}_{u}}\}=0,\{{{\mathcal{Q}}^{\dag}_{d}},{{\mathcal{Q}}^{\dag}_{d}}\}=0,{\,}\{{{\mathcal{Q}}^{\dag}_{d}},{{\mathcal{Q}}^{\dag}_{u}}\}=0,{\,}\{{{\mathcal{Q}}_{d}},{{\mathcal{Q}}_{u}}\}=0{\,}\\
\notag
&[{{\mathcal{Q}}_{u}},{{\mathcal{Q}}_{d}}]=0,[{{\mathcal{Q}}^{\dag}_{d}},{{\mathcal{Q}}^{\dag}_{u}}]=0,{\,}[{{\mathcal{Q}}_{u}},{{\mathcal{Q}}_{u}}]=0,{\,}[{{\mathcal{Q}}^{\dag}_{u}},{{\mathcal{Q}}^{\dag}_{u}}]=0
\end{align}
These relations describe a $N=4$, $d=1$ superalgebra with four non-trivial topological charges $\mathcal{Z}_{uu},\mathcal{Z}_{dd},\mathcal{Z}_{ud},\mathcal{Z}_{du}$. The next $N=4$ superalgebra is described by the following relations:
\begin{align}\label{commutatorsantcvicomm123}
&\{{{\mathcal{Q}}_{u}},{{\mathcal{Q}}^{\dag}_{u}}\}=2\mathcal{H}+\mathcal{Z}_{uu},
{\,}\{{{\mathcal{Q}}_{e}},{{\mathcal{Q}}^{\dag}_{e}}\}=2\mathcal{H}+\mathcal{Z}_{ee}
,{\,}\{{{\mathcal{Q}}_{e}},{{\mathcal{Q}}_{e}}\}=0,
\\ \notag &\{{{\mathcal{Q}}_{u}},{{\mathcal{Q}}_{ u}}\}=0, {\,} \{{{\mathcal{Q}}_{e}},{{\mathcal{Q}}_{ u}}^{\dag}\}=\mathcal{Z}_{e u},{\,}\{{{\mathcal{Q}}_{u}},{{\mathcal{Q}}^{\dag}_{e}}\}=\mathcal{Z}_{u e},{\,}\\ \notag
&\{{{\mathcal{Q}}^{\dag}_{u}},{{\mathcal{Q}}^{\dag}_{u}}\}=0,\{{{\mathcal{Q}}^{\dag}_{e}},{{\mathcal{Q}}^{\dag}_{e}}\}=0,{\,}\{{{\mathcal{Q}}^{\dag}_{e}},{{\mathcal{Q}}^{\dag}_{u}}\}=0,{\,}\{{{\mathcal{Q}}_{e}},{{\mathcal{Q}}_{u}}\}=0{\,}\\
\notag
&[{{\mathcal{Q}}_{u}},{{\mathcal{Q}}_{e}}]=0,[{{\mathcal{Q}}^{\dag}_{e}},{{\mathcal{Q}}^{\dag}_{u}}]=0,{\,}[{{\mathcal{Q}}_{u}},{{\mathcal{Q}}_{u}}]=0,{\,}[{{\mathcal{Q}}^{\dag}_{u}},{{\mathcal{Q}}^{\dag}_{u}}]=0
\end{align}
The supercharges of this algebra are $\mathcal{Q}_u,\mathcal{Q}_d$ and the topological charges are $\mathcal{Z}_{uu},\mathcal{Z}_{ee},\mathcal{Z}_{ue},\mathcal{Z}_{eu}$. Finally, the last $N=4$ superalgebra is composed by the supercharges $\mathcal{Q}_e,\mathcal{Q}_d$, which satisfy the following relations:
\begin{align}\label{commutatorsanticomm456df}
&\{{{\mathcal{Q}}_{e}},{{\mathcal{Q}}^{\dag}_{e}}\}=2\mathcal{H}+\mathcal{Z}_{ee},
{\,}\{{{\mathcal{Q}}_{d}},{{\mathcal{Q}}^{\dag}_{d}}\}=2\mathcal{H}+\mathcal{Z}_{dd}
,{\,}\{{{\mathcal{Q}}_{d}},{{\mathcal{Q}}_{d}}\}=0,
\\ \notag &\{{{\mathcal{Q}}_{e}},{{\mathcal{Q}}_{ e}}\}=0, {\,} \{{{\mathcal{Q}}_{d}},{{\mathcal{Q}}_{ e}}^{\dag}\}=\mathcal{Z}_{d e},{\,}\{{{\mathcal{Q}}_{e}},{{\mathcal{Q}}^{\dag}_{d}}\}=\mathcal{Z}_{e d},{\,}\\ \notag
&\{{{\mathcal{Q}}^{\dag}_{e}},{{\mathcal{Q}}^{\dag}_{e}}\}=0,\{{{\mathcal{Q}}^{\dag}_{d}},{{\mathcal{Q}}^{\dag}_{d}}\}=0,{\,}\{{{\mathcal{Q}}^{\dag}_{d}},{{\mathcal{Q}}^{\dag}_{e}}\}=0,{\,}\{{{\mathcal{Q}}_{d}},{{\mathcal{Q}}_{e}}\}=0{\,}\\
\notag
&[{{\mathcal{Q}}_{e}},{{\mathcal{Q}}_{d}}]=0,[{{\mathcal{Q}}^{\dag}_{d}},{{\mathcal{Q}}^{\dag}_{e}}]=0,{\,}[{{\mathcal{Q}}_{e}},{{\mathcal{Q}}_{e}}]=0,{\,}[{{\mathcal{Q}}^{\dag}_{e}},{{\mathcal{Q}}^{\dag}_{e}}]=0
\end{align}
The topological charges for this algebra are $\mathcal{Z}_{dd},\mathcal{Z}_{ee},\mathcal{Z}_{de},\mathcal{Z}_{ed}$. We summarize the results in table 1, where we have included the supercharges and the corresponding topological charges for each $N=4$ superalgebra.
\begin{center} 
    \begin{tabular}{ | p{2cm} | p{3.5cm}  | p{4cm} |}
    \hline
    $d=1$ SUSY & Supercharges that define the Algebra & Topological Charges of the Algebra\\ \hline
    1st $N=4$ Algebra & $\mathcal{Q}_{u},{\,}\mathcal{Q}_{d}$ & $\begin{array}{c}
 \mathcal{Z}_{uu},\mathcal{Z}_{ud} \\
  \mathcal{Z}_{du},\mathcal{Z}_{dd} \\
\end{array}$
     \\ \hline
    2nd $N=4$ Algebra & $\mathcal{Q}_{u},{\,}\mathcal{Q}_{e}$ & $\begin{array}{c}
 \mathcal{Z}_{uu},\mathcal{Z}_{ue} \\
  \mathcal{Z}_{eu},\mathcal{Z}_{ee} \\
\end{array}$
\\ \hline
    3rd $N=4$ Algebra & $\mathcal{Q}_{d},{\,}\mathcal{Q}_{e}$ &  $\begin{array}{c}
 \mathcal{Z}_{dd},\mathcal{Z}_{de} \\
  \mathcal{Z}_{ed},\mathcal{Z}_{ee} \\
\end{array}$ \\
    \hline
    \end{tabular}
    \\
    \bigskip 
    Table 1
\end{center}

\section{Duality Transformations and Their Impact on the Supersymmetry Algebras}

The free parameters of the grand unified domain wall model we are using in this article are the coupling constants $g_e,g_d,g_u$ and the functions $\phi_d^{kink}(y)=\phi_e^{kink}(y),\phi_u^{kink}(y)$, which show the way the domain wall couples to the fermions. If we make certain transformations on the $N=4$ and $N=6$ algebras we found in the previous sections, some of the algebras stay invariant and others transform to other algebras. The purpose of this section is to find all possible transformations of the supersymmetric algebras.

Let us start with each $N=4$, $d=1$ supersymmetric algebra and particularly with the $N=4$ that contains the supercharges $\mathcal{Q}_u,\mathcal{Q}_d$. The following set of transformations:
\begin{align}\label{fset}
&g_u\phi_u^{kink}(y)\rightarrow g_d\phi_d^{kink}(y) \\ \notag &
g_d\phi_d^{kink}(y)\rightarrow g_u\phi_u^{kink}(y)
\end{align} 
leave the algebra invariant, since the transformations (\ref{fset}) are equivalent to the transformations
\begin{align}\label{gfdfg}
 &\mathcal{Q}_u\rightarrow \mathcal{Q}_d \\ \notag &
 \mathcal{Q}_d\rightarrow \mathcal{Q}_u
\end{align}
Accordingly, the algebra that contains the supercharges $\mathcal{Q}_u,\mathcal{Q}_e$, is invariant under the transformations:
\begin{align}\label{fset1}
&g_u\phi_u^{kink}(y)\rightarrow g_e\phi_e^{kink}(y) \\ \notag &
g_e\phi_e^{kink}(y)\rightarrow g_u\phi_u^{kink}(y)
\end{align} 
since the transformations (\ref{fset1}) are equivalent to the transformation:
\begin{align}\label{gfdfg1}
&\mathcal{Q}_u\rightarrow \mathcal{Q}_e \\ \notag &
\mathcal{Q}_e\rightarrow \mathcal{Q}_u
\end{align}
In the same vain, the algebra described by $\mathcal{Q}_d,\mathcal{Q}_e$, is invariant under the simple transformation:
\begin{align}\label{fset11}
&g_e\rightarrow g_d \\ \notag &
g_d\rightarrow g_e
\end{align} 
since $\phi_e^{kink}(y)=\phi_d^{kink}(y)$ and the transformations (\ref{fset11}) are equivalent to the transformation:
\begin{align}\label{gfdfg111}
&\mathcal{Q}_d\rightarrow \mathcal{Q}_e \\ \notag &
\mathcal{Q}_e\rightarrow \mathcal{Q}_d
\end{align}
Apart from the transformations (\ref{fset}), (\ref{fset1}) and (\ref{fset11}), there exist other transformations that transform one $N=4$ algebra to another. We start with the $N=4$ algebra described by the supercharges $\mathcal{Q}_u,\mathcal{Q}_e$ and upon making the substitution,
\begin{equation}\label{sub1}
g_u\phi_u^{kink}(y)\rightarrow g_d\phi_d^{kink}(y)
\end{equation}
the algebra is transformed to another $N=4$ algebra described by the supercharges $\mathcal{Q}_d,\mathcal{Q}_e$. In the same way, by making the substitution, 
\begin{equation}\label{sub11}
g_d\rightarrow g_e
\end{equation}
the algebra with supercharges $\mathcal{Q}_u,\mathcal{Q}_d$ is transformed to one with supercharges $\mathcal{Q}_u,\mathcal{Q}_e$ and so on. 

Now we turn our focus to the $N=6$, $d=1$ supersymmetric algebra which is invariant under the set of transformations:
\begin{align}\label{fset1n6total}
&g_u\phi_u^{kink}(y)\rightarrow g_d\phi_d^{kink}(y) \\ \notag &
g_d \rightarrow g_e \\ \notag &
g_e\phi_e^{kink}(y)\rightarrow g_u\phi_u^{kink}(y)
\end{align} 
since $\phi_e^{kink}(y)=\phi_d^{kink}(y)$ and also the transformations (\ref{fset1n6total}) are equivalent to the following set of transformations:
\begin{align}\label{gfdfg111}
&\mathcal{Q}_u\rightarrow \mathcal{Q}_d \\ \notag &
\mathcal{Q}_d\rightarrow \mathcal{Q}_e \\ \notag &
\mathcal{Q}_e\rightarrow \mathcal{Q}_u
\end{align}
However, if we apply only one of the three transformations appearing in (\ref{fset1n6total}), to the $N=6$ algebra, say $g_u\phi_u^{kink}(y)\rightarrow g_d\phi_d^{kink}(y)$, the original $N=6$ algebra that contains the three supercharges $\mathcal{Q}_u,\mathcal{Q}_d,\mathcal{Q}_e$, breaks to one $N=4$ algebra with supercharges $\mathcal{Q}_d,\mathcal{Q}_e$. The same applies if we take into account only one the rest two transformations appearing in (\ref{fset1n6total}), and therefore, the initial $N=6$, $d=1$ supersymmetry explicitly breaks to a single $N=4$, $d=1$.

\section*{Conclusions}

In this letter we demonstrated that the fermionic system of zero modes on a superconducting domain wall has a rich extended one dimensional supersymmetric structure. Particularly, the supersymmetry we found in this letter is a $N=6$, $d=1$ supersymmetry with non-trivial topological charges. In addition, the system also has three distinct $N=4$, $d=1$ supersymmetries, with non-trivial topological charges, as can be seen in table 2 where we gathered our results.
\begin{center} 
    \begin{tabular}{ | p{3cm} | p{3cm}  | p{3cm}  |p{3cm}  |}
    \hline
    Supersymmetric Structure & Number of Different Supersymmetries & Number of Supercharges of Each Symmetry & Number of Topological Charge of Each Symmetry \\ \hline
    $N=4$ & $3$ & 2 & 4
     \\ \hline
    $N=6$ & 1 & 3 & 10
\\ \hline
    
    \end{tabular}
    \\
    \bigskip 
    Table 2
\end{center}
In addition, we studied some dualities of the one dimensional superalgebras we found. The present work is a generalization of a previous work \cite{oikonomoudomain}, where we found that a $N=4$, $d=1$ supersymmetry underlies the fermionic system that is composed by the lepton and down quark, in the particular case of gauge coupling unification. We believe that the work we presented in the present article generalizes the work of the previous one and also motivates some questions, with regards the existence of non-trivial topological charges. As we already mentioned in the text, the presence of non-trivial topological charges that do not commute with the operators of the superalgebra, indicates two possible things. Firstly, that the extended supersymmetry is not a symmetry of the $S$-matrix of the theory (obviously) and secondly the fact that there exists a possible underlying non-linear supersymmetric structure yet to be found. In reference to this non-linear supersymmetry, it is known in the literature \cite{rossi} that the very own existence of fermionic zero modes in the presence of a topological defect is related to a hidden underlying supersymmetry of the fermionic system. The result found in \cite{rossi} is different in spirit from problems of fermions and their localization properties on higher dimensional defects like branes (see \cite{liu} and references therein). Therefore, it would be interesting to connect directly the existence of fermion zero modes on the domain wall with non-linear supersymmetry, a task that exceeds the purpose of the present work.

Before closing, we briefly discuss another important issue, having to do with $d=4$ global supersymmetry and $d=1$ supersymmetry. Although the initial fermionic system on the domain wall had no global supersymmetry of any kind, we revealed a rich non-trivial supersymmetric underlying structure that can get even more involved when the number of flavors increases. This behavior has been pointed out in the literature before \cite{oiko1}. The present findings motivate us to conclude that global supersymmetry plays no obvious role in the form of the one dimensional supersymmetries we found in this article, since the latter are symmetries of the field theory and not of the $S$-matrix.


\begin{thebibliography}{99}

\bibitem{vilenkin} A. Vilenkin, E. P. S. Shellard, Cosmic Strings and Other Topological Defects, Cambridge University Press (2000) 

\bibitem{lazaridesvasiko} G. Lazarides, Q. Shafi, Phys. Lett. B159 (1985) 261; T. W. B. Kibble, G. Lazarides, Q. Shafi, Phys. Rev. D26 (1982) 435; T. W. B. Kibble, G. Lazarides, Q. Shafi, Phys. Lett. B113 (1982) 237; J. R. Morris, Phys. Rev D52 (1995) 1096; D. Stojkovic, Phys. Rev. D63 (2001) 025010; Dejan Stojkovic, Int.J.Mod.Phys.A16 (2001) 1034

\bibitem{odintsovgravity}  S. Nojiri, S. D. Odintsov, Int. J. Geom. Meth. Mod. Phys. 4 (2007) 115; S. Nojiri, S. D. Odintsov, Phys. Rept. 505 (2011) 59; G. Cognola, E. Elizalde, S. Nojiri, S.D. Odintsov, L. Sebastiani, S. Zerbini, Phys.Rev. D77 (2008) 046009;  A. V. Astashenok, S. Capozziello, S. D. Odintsov, arXiv:1309.1978; H. Farajollahi, J. Sadeghi, M. Pourali, Astrophys.Space Sci. 341 (2012) 695; M. Khurshudyan, E. O Kahya, A. Pasqua, B. Pourhassan, arXiv:1401.6630


\bibitem{oikonomoudomain} V. K. Oikonomou, Class. Quant. Grav. 31 (2014) 025018


\bibitem{reviewsusyqm} M. de Crombrugghe, V. Rittenberg, Annals Phys. 151 (1983) 99; C.V. Sukumar, J.Phys. A18 (1985) 2917;
M. de Crombrugghe, V. Rittenberg
Annals Phys. 151 (1983) 99; F. Cooper, B. Freedman, Annals Phys. 146 (1983) 262; M. Henneaux, C. Teitelboim,
Annals Phys. 143 (1982) 127; F. Cooper, A. Khare, U. Sukhatme, Phys.Rept. 251 (1995) 267; F. Cooper, J. N. Ginocchio, Phys. Rev. D 36, 2458 (1987);  G. Junker, ``Supersymmetric Methods in Quantum and Statistical Physics", Springer, 1996


\bibitem{diffgeomsusyduyalities}

 D. Ruan, C.C. Tu, H.Z. Sun, Commun.Theor.Phys. 32 (1999) 477; D. Spector, Int.J.Mod.Phys. A20 (2005) 6288;  C. Quesne, Mod.Phys.Lett. A18 (2003) 515


\bibitem{susyqminquantumsystems}

B. Bagchi, S. Mallik, C. Quesne, Int. J. Mod. Phys. A16 (2001) 2859;  V. I. Tkach, Pashnev A. I., Rosales J. J., Mod.Phys.Lett. A15 (2000) 1557


\bibitem{plu1}  M. S. Plyushchay, A. Wipf, arXiv:1311.2195; A. Arancibia, M. S. Plyushchay, arXiv:1401.6709; A. Arancibia, J. M. Guilarte, M. S. Plyushchay, Phys. Rev. D88 (2013) 085034; A. Arancibia, J. M. Guilarte, M. S. Plyushchay, Phys. Rev. D87 (2013) 4, 045009; A. A. Izquierdo, J. M. Guilarte, M. S. Plyushchay, Annals Phys. 331 (2013) 269; M. S. Plyushchay, A. Arancibia, L. M. Nieto, Phys.Rev. D83 (2011) 065025; A. Arancibia, M. S. Plyushchay, Phys.Rev. D85 (2012) 045018; F. Correa, V. Jakubsky, L. M. Nieto, Mikhail S. Plyushchay, Phys.Rev.Lett. 101 (2008) 030403; F. Correa, V. Jakubsky, M. S. Plyushchay, J.Phys. A41 (2008) 485303; F. Correa, V. Jakubsky, M. S. Plyushchay, Annals Phys. 324 (2009) 1078;


\bibitem{plu2}F. Correa, L. M. Nieto, M. S. Plyushchay, Phys.Lett. B659 (2008) 746


\bibitem{susyqmscatter}

V.P. Berezovoi, A.I. Pashnev, Theor.Math.Phys. 74 (1988) 264, Teor.Mat.Fiz. 74 (1988) 392

\bibitem{susybreaking} E. Witten, Int.J.Mod.Phys.A10:1247 (1995)

\bibitem{ivanov} F. Delduc, E.A. Ivanov, Nucl. Phys. B
855 (2012) 815; F. Delduc, S. Kalitzin, E. Sokatchev, Class. Quantum Grav. 7 (1990) 1567; E. Ivanov, O. Lechtenfeld, A. Sutulin,  Nucl. Phys.
B 790 (2008) 493

\bibitem{extendedsusy}

Z. Kuznetsova , F. Toppan, Int. J. Mod. Phys. A23 (2008) 3947

\bibitem{witten1} E. Witten, Nucl. Phys. B188 (1981) 513 

\bibitem{odi1} Kiwoon Choi, Adam Falkowski, H. P. Nilles, M. Olechowski, Nucl.Phys. B718 (2005) 113; S.D. Odintsov, I.L. Shapiro, Mod.Phys.Lett. A4 (1989) 1479; Ann E. Nelson, N. Seiberg, Nucl.Phys. B416 (1994) 46

\bibitem{odi2} I.L. Buchbinder, S.D. Odintsov, Int.J.Mod.Phys. A4 (1989) 4337; K. A. Intriligator, S. D. Thomas, Nucl.Phys. B473 (1996) 121

\bibitem{odi3} F. Aceves de la Cruz, J.J. Rosales, V.I. Tkach, J. Torres A, Grav.Cosmol. 8 (2002) 101; S.D. Odintsov, Mod.Phys.Lett. A3 (1988) 1391

\bibitem{haag} Rudolf Haag, Jan T. Lopuszanski, Martin Sohnius, Nucl. Phys. B88 (1975) 257

\bibitem{wittentplc} E. Wittem, D. Olive, Phys.Lett. B78 (1978) 97

\bibitem{fayet} P. Fayet, Nucl. Phys. B149 (1979) 137

\bibitem{topologicalcharges} K. Shizuya, Phys. Rev. D70 (2004) 065003

\bibitem{topologicalcharges1} I. Gaida, Phys.Lett. B373 (1996) 89

\bibitem{topologicalcharges2} K. Kamimura, M. Sakaguchi,  Nucl.Phys. B662 (2003) 491

\bibitem{oiko1} V. K. Oikonomou, Nucl. Phys. B870 (2013) 477

\bibitem{rossi} P. Rossi, Phys.Lett. B71 (1977) 145

\bibitem{liu} X. N. Zhou, X. L. Du, K. Yang, Yu-Xiao Liu, arXiv:1308.2863; Yu-Xiao Liu, L. Zhao, X. H. Zhang, Y. S. Duan, Nucl.Phys. B785 (2007) 234;  J. Sadeghi, B. Pourhassan, K. Jafarzadeh, E. Reisi, M. Rostami, Can.J.Phys. 91 (2013) 251






\end{thebibliography}
\end{document}